\documentclass[prl,twocolumn,showpacs,superscriptaddress,floatfix]{revtex4}

\usepackage{graphicx}
\usepackage{dcolumn} 
\usepackage{amsmath, amsthm, amssymb, amsfonts, bm}

\begin{document}
\title{Creation and detection of skyrmions in a Bose-Einstein condensate}

\author{L. S. Leslie}
\affiliation{The Institute of Optics, University of Rochester, Rochester, NY 14627}
         
\author{A. Hansen}
\affiliation{Department of Physics and Astronomy, University of Rochester, Rochester, NY 14627}

\author{K. C. Wright$^\dagger$}
\affiliation{Department of Physics and Astronomy, University of Rochester, Rochester, NY 14627}

\author{B. M. Deutsch}
\affiliation{The Institute of Optics, University of Rochester, Rochester, NY 14627}

\author{N. P. Bigelow}
\affiliation{Department of Physics and Astronomy, University of Rochester, Rochester, NY 14627}
\affiliation{The Institute of Optics and Laboratory for Laser Energetics, University of Rochester, Rochester, NY 14623}
\date{\today}  

\begin{abstract}
We present the first experimental realization and characterization of two-dimensional skyrmions and half-skyrmions in a spin-2 Bose-Einstein condensate.  The continuous rotation of the local spin of the skyrmion through an angle of $\pi$ (and half-skyrmion through an angle of $\pi/2$) across the cloud is confirmed by the spatial distribution of the three spin states as parameterized by the bending angle of the $\ell$-vector.  The winding number, $w=\left(0,1,2\right)$, of the skyrmions is confirmed through matter-wave interference.
\end{abstract}

\pacs{03.75.Mn, 03.75.Lm, 37.25.+k,37.10.Vz}
\maketitle

Skyrmions are topological solitons that were first envisioned in the 1960's as part of a nonlinear field theory to model mesons and baryons in nuclear physics \cite{SkyrmeNon-linearPRSLA61}.  They are one of the many topological defects that can be described and classified by homotopy groups of their order-parameter space \cite{MerminTopologicalRMP79,MakelaTopologicalJPA03}.  A 3D skyrmion is an $S^3 \rightarrow S^3$ map that can be labeled by an integer representing the topological degree, or winding number, of the map \cite{RuostekoskiCreatingPRL01}.  The 2D or ``baby skyrmions'' are characterized by a local spin that continuously rotates through an angle of $\pi$ from the center to the boundary of the system \cite{MantonTopological04}, and are the focus of this Letter.  The search for skyrmions spans several fields including superfluids, solid state physics, liquid crystals, and superconductors \cite{AndersonPhasePRL77,MerminCirculationPRL76,KoralekEmergenceNL09,RosslerSpontaneousN06,BogdanovSkyrmionsPRE03,KevrekidisSkyrmion-likePRE07,PereiraPseudo-particlesJPCM07}.  The experimental results presented in this Letter mark the first deterministic creation of skyrmions and half-skyrmions in a spin-2 Bose-Einstein condensate (BEC).  

As in many condensed matter systems, a BEC of alkali metal atoms can be described by an order parameter which becomes vectorial with $2F+1$ components when multiple states in the same hyperfine spin-$F$ manifold are simultaneously populated \cite{HoSpinorPRL98}.  These spinor condensates have made it possible to study experimentally a multitude of spin excitations such as coreless vortices, solitons, and spin domains in a dilute system instead of one dominated by the strong interactions and high densities of condensed matter physics \cite{LeanhardtCorelessPRL03,MatthewsVorticesPRL99,StengerSpinNature98,LeggettBose-EinsteinRMP01}.  

For a spin-2 condensate, two five-component order parameters are required to sufficiently define the system: $\left\langle f \right\rangle$, which describes the ferromagnetic character, and $\Theta$, which is related to the dynamic creation of spin singlets through collisions \cite{CiobanuPhasePRA00,PogosovVortexPRA05}.  In our experiments, prior to the creation of the skyrmion, the BEC is spin-polarized in the $\left|F=2,m_F=2\right\rangle=\left|2\right\rangle$ state with a density of $\approx 10^{11}$ cm$^{-3}$.  This means that initially $\left\langle f \right\rangle=2$, while $\Theta$ is negligible for the timescales involved. 

A detailed description of our coherent Raman interaction can be found in \cite{WrightRamanPRA08}.  At the start of the experiment, the BEC is untrapped and cylindrically symmetric with a magnetic field of $B=1.33$ G oriented along the $z$-axis.  Two beams ($\sigma^-$, $\sigma^+$ polarized) propagate collinearly and parallel to the quantization axis, creating a double $\Lambda$ (or ``M'') system by simultaneously coupling the $\left|2\right\rangle$ state with the $\left|2,0\right\rangle=\left|0\right\rangle$ and $\left|2,-2\right\rangle=\left|-2\right\rangle$ ground states.  The Raman beams are applied diabatically to the BEC in $5$ $\mu$s pulses.  The short timescale enables the creation of non-equilibrium spin textures in the condensate, while the co-propagating beam geometry minimizes the transfer of linear momentum.    

The $\sigma^-$, $\sigma^+$ Raman beams have first-order Laguerre-Gaussian (LG$^{-1}$) and Gaussian intensity profiles, respectively, so the population transferred to the $\left|0\right\rangle$ ($\left|-2\right\rangle$) state acquires a $w=1$ ($w=2$) azimuthal phase winding.  A core of $w=0$ atoms is left in the initial $\left|2\right\rangle$ state corresponding to the intensity minimum of the LG beam.  The relative population transfer to the $\left|0\right\rangle$, $\left|-2\right\rangle$ states is a function of the two-photon detuning of the Raman beams \cite{WrightRamanPRA08}, while the spatial dependence of the three components of the coreless vortex additionally depends on the beamwaist of the LG beam.  It is in this way that the Raman beam coupling, combined with the relative energetic degeneracy of the $\left|2\right\rangle \leftrightarrow \left|0\right\rangle$ and $\left|0\right\rangle\leftrightarrow\left|-2\right\rangle$ transitions, creates a coreless vortex with winding number: $w=(0,\times,1,\times,2)$.  The $\times$ indicates that there is no population in a spin state, and so there can be no phase winding there (the $\times$ in the $\left|2,1\right\rangle$, $\left|2,-1\right\rangle$ states will be dropped).  The Raman interaction effectively evolves the order parameter of the initially spin-polarized BEC to
\begin{equation}\label{eq:spinor}
\Psi(r)=\sqrt{n(r)}
\left( \begin{array}{ccc} 
\cos^2(\beta(r)/2)         
\\
0
\\
\sqrt{2} e^{i\varphi} \sin(\beta(r)/2)\cos(\beta(r)/2)
\\
0
\\
e^{2i\varphi}\sin^2(\beta(r)/2)
\end{array}\right).
\end{equation} 

\noindent Here $n(r)$ is the density of the cloud, ($r$, $\varphi$) are polar coordinates, and $\beta(r)$ is the bending angle which characterizes the rotation or ``bending'' of the local spin across the cloud.  The local spin is represented by the spin texture's $\ell$-vector: 
\begin{equation}\label{eq:lvect}
\vec{\ell}(r,\varphi)=\hat{z}\cos\beta(r)+ \sin\beta(r)(\hat{x}\cos\varphi+ \hat{y}\sin\varphi).  
\end{equation}

\noindent This $\ell$-vector description of the local spin is a general definition of a 2D skyrmion, and it requires that the bending angle be monotonic and satisfy $\beta(0)=0$ and $\beta(R)=\pi$, where $R$ is the boundary of the cloud \cite{MantonTopological04}.  For the half-skyrmion, the $\ell$-vector sweeps through half the angle of the skyrmion, and so $\beta$ must monotonically satisfy $\beta(0)=0$, $\beta(R)=\pi/2$ \cite{MerminCirculationPRL76,MizushimaMermin-HoPRL02}.  The continuous $\ell$-vector and non-singular order parameter are fundamental properties of skyrmions and will be discussed in more depth later.    

\begin{figure}[ht]
	\includegraphics[scale = 1.0]{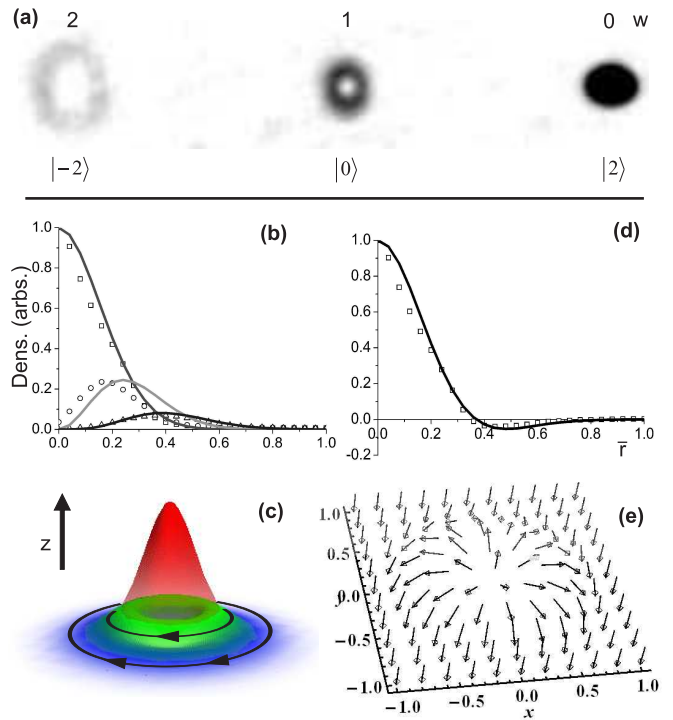}
	\caption{The absorption image (a) of a 2D skyrmion created in spin-2 $^{87}$Rb. The winding number, $w$, for each spin state is indicated. The contrast of the image has been increased to make the $\left|-2\right\rangle$ state more visible.  The relative populations of the $\left|2\right\rangle$, $\left|0\right\rangle$, and $\left|-2\right\rangle$ states are: $51\%, 30\%, 19\%$. (b) Azimuthally averaged lineouts (points) for each spin state agree well with the plots of (\ref{eq:spinor}) (solid lines) for the $\beta_1(\bar{r})$ determined by a nonlinear least-squares regression. (c) 3D plot of the solid lines in (b), where the colors red, green, and blue correspond to the $\left|2\right\rangle$, $\left|0\right\rangle$, and $\left|-2\right\rangle$ states, respectively. The winding number of the spin state is indicated by the number of arrowheads. (d) The polarization of the skyrmion.  The points represent $\left|\phi_2(\bar{r})\right|^2-\left|\phi_{-2}(\bar{r})\right|^2$ from (b), and closely match the solid line: $\ell_z (\bar{r})= n(\bar{r})\cos\beta_1(\bar{r})$. (e) The vector plot of the skyrmion $\ell$-vector clearly shows the rotation of the local spin through an angle of $\pi$ across the cloud.  The same $\beta_1$ was used for all theory plots, and the goodness of fit for (b), (d) is $R^2_1 = 0.977$, $R^2_1=0.982$, respectively.}
	\label{fig:anderson-toulouse}
\end{figure}

\begin{figure}[ht]
	\includegraphics[scale = 1.0]{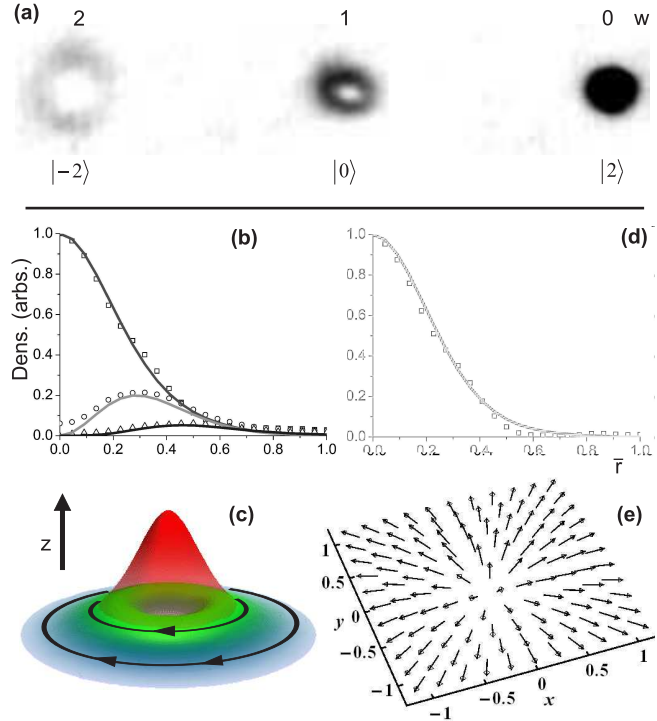}
	\caption{Absorption image (a) of a half-skyrmion created in spin-2 $^{87}$Rb.  The lineouts (b), reconstruction (c), polarization (d) and $\ell$-vector (e) have been fit, processed and presented as in Fig. \ref{fig:anderson-toulouse}.  The relative populations of the $\left|2\right\rangle$, $\left|0\right\rangle$, and $\left|-2\right\rangle$ states are: $54\%, 31\%, 15\%$, and the goodness of fit for (b), (d) is $R^2_2=0.996$, $R^2_2=0.999$, respectively.}
	\label{fig:mermin-ho}
\end{figure}

Previous experiments conducted in spin-2 $^{87}$Rb have focused on equilibrium studies and spin-mixing dynamics without vortices \cite{HallDynamicsPRL98}.  Initial coreless vortex work has involved coupling the $F=1,2$ ground state manifolds of $^{87}$Rb by RF transitions, and the adiabatic manipulation of spin-1 $^{23}$Na by magnetic fields \cite{LeanhardtCorelessPRL03,MatthewsVorticesPRL99}.  While novel in their approach, these coreless-vortex techniques are limited in their applicability: the former by its inability to create multi-component coreless vortices within a single ground state manifold, and the latter due to the inherent difficulty of creating and controlling arbitrary magnetic fields.  In contrast, we can tailor the Raman interaction used here to create $w=(0,1,2)$ coreless vortices while simultaneously controlling the spatial distributions of the spin states and the normalized magnetization of the cloud in order to precisely engineer complicated spin textures such as skyrmions (Fig. \ref{fig:anderson-toulouse}) and half-skyrmions (Fig. \ref{fig:mermin-ho}).   

The $(0,1,2)$ coreless vortex presented in Fig. \ref{fig:anderson-toulouse} is a 2D skyrmion created in a spin-2 BEC.  The absorption image, \ref{fig:anderson-toulouse}(a), shows the three components of the cloud after they have been spatially separated by an inhomogeneous magnetic field for imaging.  Prior to this ``Stern-Gerlach'' pulse however, the coreless vortex is axi-symmetric and the cloud maintains its cylindrical symmetry (Fig. \ref{fig:anderson-toulouse}(c)).  The spatial dependence of the spin states comprising the spin texture, Fig. \ref{fig:anderson-toulouse}(b), can be used to find the bending angle of the cloud by fitting (\ref{eq:spinor}) to lineouts taken from the absorption image.  The solid lines in \ref{fig:anderson-toulouse}(b) are the result of simultaneously fitting all three spin states to (\ref{eq:spinor}) for a single bending angle, $\beta_1$.  The agreement confirms that this spin texture is indeed a skyrmion.  

The polarization of the cloud, Fig. \ref{fig:anderson-toulouse}(d), is one of the most recognizable differences between skyrmions and half-skyrmions.  If the density of the cloud were constant, the polarization, or local magnetization, of the skyrmion would be equivalent to the $z$-component of the $\ell$-vector: $\ell_z=\cos\beta$ \cite{MizushimaMermin-HoPRL02}.  Since this is not the case for a BEC, the plot of $\ell_z$ has been multiplied by the density profile of the cloud.  This necessarily supresses the amplitude of the polarization at the boundary, but in the outer region of the cloud the dominance of the $\left|-2\right\rangle$ state still causes the polarization to become negative.  It should be noted that the solid line is not a fit to the polarization data, instead it is a curve created using the bending angle generated by fitting (\ref{eq:spinor}) to the absorption image.  The close agreement confirms that $\beta_1$ parameterizes the cloud well, and can therefore be used to reveal the spatial dependence of the local spin of the cloud.  

Fig. \ref{fig:anderson-toulouse}(e) presents the $\ell$-vector of the skyrmion: initially parallel to the $z$-axis at the center of the cloud, it continuously rotates through an angle of $\pi$ to lie anti-parallel to the $z$-axis at $\bar{r}=r/R=1$.  This is a cylindrically symmetric 2D spin texture, analogous to the Anderson-Toulouse spin texture predicted to exist in superfluid helium \cite{AndersonPhasePRL77}.  In contrast to the skyrmion ground states predicted to arise spontaneously in magnetic metals \cite{RosslerSpontaneousN06}, the rotation of the $\ell$-vector is in the $r-z$ plane.  This spin texture is reminiscent of the spin helices recently observed in semiconductor quantum wells \cite{KoralekEmergenceNL09}, and those predicted to occur in nematic liquid crystals with phase angle $\alpha = 0$ \cite{BogdanovSkyrmionsPRE03}.  Creating a skyrmion in a BEC requires a coreless vortex where the successive peaks in the spatial distributions of the components contribute to the continuous bending of the $\ell$-vector, while simultaneously ensuring that the order parameter is non-singular.    

The half-skyrmion or meron spin texture was first envisioned in the 1970's, and was predicted to form spontaneously in superfluid helium \cite{MerminCirculationPRL76,MizushimaMermin-HoPRL02,AffleckMassPRL86}.  Fig. \ref{fig:mermin-ho} shows the first confirmed creation of a half-skyrmion.  This 2D spin texture can also be described by (\ref{eq:spinor}), but the bending angle must monotonically approach $\pi/2$ at the boundary of the cloud instead of $\pi$.  This has a profound effect on the spatial distribution of the three components of the coreless vortex, as can be seen in \ref{fig:mermin-ho}(b,c) and the resulting spin texture, (e).  

The polarization, \ref{fig:mermin-ho}(d), of the half-skyrmion is qualitatively distinct from that of the skyrmion.  Maximum at the origin, it monotonically tends to zero across the cloud.  This requires $\left|\phi_2(r)\right|^2$ and $\left|\phi_{-2}(r)\right|^2$ to decay to zero together as $r \rightarrow R$, and places an upper bound on the relative population of the $\left|-2\right\rangle$ state.  These features are clearly present in \ref{fig:mermin-ho}(b) and particularly (c), where the presence of all three spin states (represented by red, green, and blue) at the boundary of the cloud results in a turquoise color instead of the solid blue (indicative of the $\left|-2\right\rangle$ state) of the skyrmion (Fig. \ref{fig:anderson-toulouse}(c)).  

By fitting (\ref{eq:spinor}) to the absorption image lineout data in Fig. \ref{fig:mermin-ho}(b), we again find a bending angle that parameterizes the spatial dependence of the spin states.  The resulting $\ell$-vector of the half-skyrmion is shown in Fig. \ref{fig:mermin-ho}(e).  The local spin of the cloud still points along $z$ at $\bar{r}=0$, but in contrast to the skyrmion, rotates only through an angle of $\pi/2$ to lie in the $x-y$ plane at the boundary.  The effect of the monotonicity constraint on the bending angle parameter is more obvious for the half-skyrmion.  At any point where $\beta > \pi/2$ the local spin would be negative, violating the definition of a half-skyrmion.  Recent theoretical work in high temperature superconductivity has featured the half-skyrmion \cite{PereiraPseudo-particlesJPCM07}, but it has yet to be observed experimentally in those systems.      

\begin{figure}[ht]
	\includegraphics[scale = 1.0]{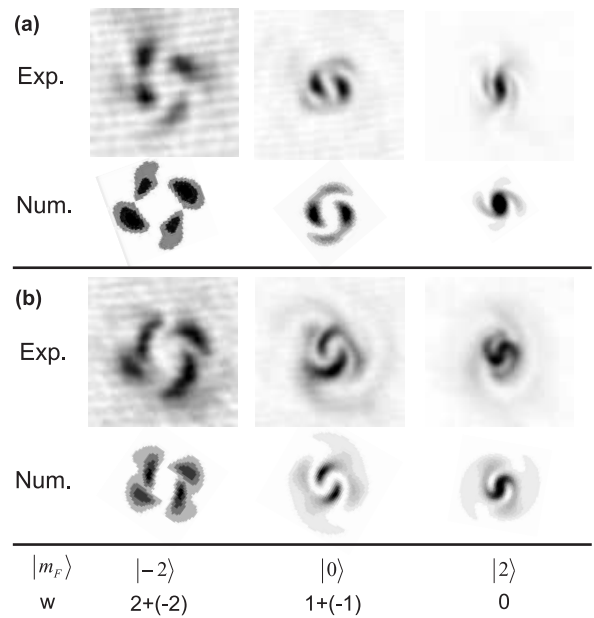}
	\caption{Matter-wave interference of $(0,1,2)$ and $(0,-1,-2)$ coreless vortices alongside numerical predictions.  The contrast of the data and corresponding theory has been increased for the $\left|0\right\rangle$, $\left|-2\right\rangle$ states to make the structure more visible.  The relative populations of the $\left|2\right\rangle$, $\left|0\right\rangle$, $\left|-2\right\rangle$ states for (a) and (b) are: $49\%, 30\%, 21\%$ and $50\%, 30\%, 20\%$.}
	\label{fig:interference}
\end{figure}

The azimuthal phase winding of the spin states comprising the skyrmions is critical to their topological stability.  We use matter-wave interference to directly confirm that the Raman interaction is creating the $(0,1,2)$ coreless vortex structure.  The diabatic pulses and non-equilibrium nature of the interaction make it possible to generate coherent superpositions of coreless vortex states through repeated application of the Raman beams \cite{KapaleVortexPRL05}.  By changing the order of the $\sigma^-$ beam from LG$^{-1}$ to LG$^{+1}$ between the pulses, we create a superposition of a $(0,1,2)$ and $(0,-1,-2)$ coreless vortex.  The interference patterns revealed upon absorption imaging confirm the vorticity of the spin states \cite{WrightSculptingPRL09}.  Fig. \ref{fig:interference} (a) and (b) show absorption images of such superpositions, created under experimental conditions simliar to the skyrmion and half-skyrmion presented in Figs. \ref{fig:anderson-toulouse} and \ref{fig:mermin-ho}, respectively.  The two azimuthal nodes in the $\left|0\right\rangle$ state confirm that the interfering clouds have orthogonal azimuthal phase windings: $1-(-1)=2$, while the cloverleaf interference pattern in the $\left|-2\right\rangle$ state is clear evidence of the superposition of a $w=+2$, $w=-2$ vortex state.  This simultaneously demonstrates the stability of these spin textures over the $20$ ms of time-of-flight expansion between their creation and imaging.   

Skyrmions can be created across a range of magnetizations.  The normalized magnetization is critical in determining the stability and subsequent evolution of spin textures \cite{MizushimaMermin-HoPRL02,PogosovVortexPRA05,TakahashiVortex-SplittingPRA09}, as it is conserved for many interactions \cite{HallDynamicsPRL98}.  For a spin-2 system, $M/N$ ranges from $[-2, 2]$, and the skyrmions presented in Figs. \ref{fig:anderson-toulouse}, \ref{fig:mermin-ho} have normalized magnetizations of $M/N=0.64$ and $0.78$, respectively.  It is unknown at present if the ground state phase in zero magnetic field for spin-2 $^{87}$Rb is polar or cyclic \cite{CiobanuPhasePRA00}.  However, the range of normalized magnetizations for a $(0,1,2)$ coreless vortex to be stable in polar spin-2 $^{87}$Rb is estimated to be $[0.25,1.3]$ \cite{PogosovVortexPRA05}, and so from this perspective we would expect both of these spin textures to be stable.  The implementation of an optical dipole trap is underway to enable studies of the stability and evolution of not only the skyrmions presented here, but also other non-trivial coreless vortex states \cite{PietilaCreationPRL09}. 

In this Letter we have presented, in the form of $(0,1,2)$ coreless vortices, the first experimental realization of skyrmions and half-skyrmions in a spin-2 condensate.  The coreless vortices were shown to be well-characterized by bending angles, $\beta_{1,2}(\bar{r})$, that enabled the reconstruction of the $\ell$-vectors of the respective 2D spin textures.  This revealed not only the continuous evolution of the local spin of the clouds, but also their non-singular profiles.  Further investigation into the stability of skyrmions in spinor condensates has the potential to strengthen the connections between BEC physics and other systems with spin degrees of freedom \cite{LeonhardtHowJETPLett00}.   

This work was supported by the NSF and ARO. The authors would like to thank E. J. Mueller, T. Mizushima, and T-L Ho for useful conversations and correspondences.  LSL is grateful for a Horton Fellowship from the LLE. 

$\dagger$Current address: \textit{Atomic Physics Division, National Institute of Standards and Technology, Gaithersburg, Maryland, 20899}

\end{document}